\documentclass[runningheads,a4paper]{llncs}

\usepackage{amssymb}
\usepackage{graphicx}

\usepackage{url}
\urldef{\mailsa}\path|Adrian.Paschke@inf.fu-berlin.de|

\begin{document}

\mainmatter  

\title{Rule Responder: A Rule-Based Semantic eScience Service Infrastructure}

\titlerunning{Rule Responder: A Rule-Based Semantic eScience Service Infrastructure}

%
%
\author{Adrian Paschke\and Zhili Zhao}
\authorrunning{Rule Responder: A Rule-Based Semantic eScience Service Infrastructure}

\institute{Freie Universitaet Berlin, Germany\\
\mailsa}

%
%

\maketitle

\begin{abstract}
To a large degree information and services for chemical e-Science have become accessible -anytime, anywhere -but not necessarily useful. The Rule Responder eScience middleware is about providing information consumers with rule-based agents to transform existing information into relevant information of practical consequences, hence providing control to the end-users to express in a declarative rule-based way how to turn existing information into personally relevant information and how to react or make automated decisions on top of it.

\end{abstract}

\section{Introduction}

The Semantic Web builds upon XML as the common machine-readable syntax to structure content and data, upon RDF as a simple language to express property relationships between arbitrary resources identified by URIs, and ontology languages such as RDFS or OWL as a means to define rich vocabularies (ontologies) which are then used to precisely describe resources, their relations and their semantics. This prepares an infrastructure to share the relevant meaning of content and leads to a more machine-processable and relevant Web.

Many bioinformatics projects, such as UniProt, Tambis, FungalWeb, YeastHub, BioPax have meanwhile adopted the Semantic Web approach (in particular the RDF standard) and large ontologies such as the Gene Ontology are provided as RDFS or OWL ontologies. This has been utilized by several bioinformatics projects, such as W3C HCLS RDF or Bio2RDF, to solve the old problem of distributed heterogeneous data integration in health care and life sciences.

The goal of this article is to show how the Rule Responder approach can be used to build a flexible, loosely-coupled and service-oriented eScience infrastructure which allows wrapping the existing web data sources, services and tools by rule-based agents which access and transform the existing information into relevant information of practical consequences for the end-user.

\section{A Rule-Based Pragmatic Agent Web Model for Virtual eScience Infrastructures}

A virtual eScience infrastructure consists of a community of independent and often distributed (sub-) organizations which are typically represented by an organizational agent and a set of associated individual  agents. The organizational agent might act as a single agent towards other internal and external individual or organizational agents. In particular, a virtual organization's agent can be the single (or main) point of entry for communication with the "outer" world.

\begin{figure}
\centering
\includegraphics[height=6.2cm]{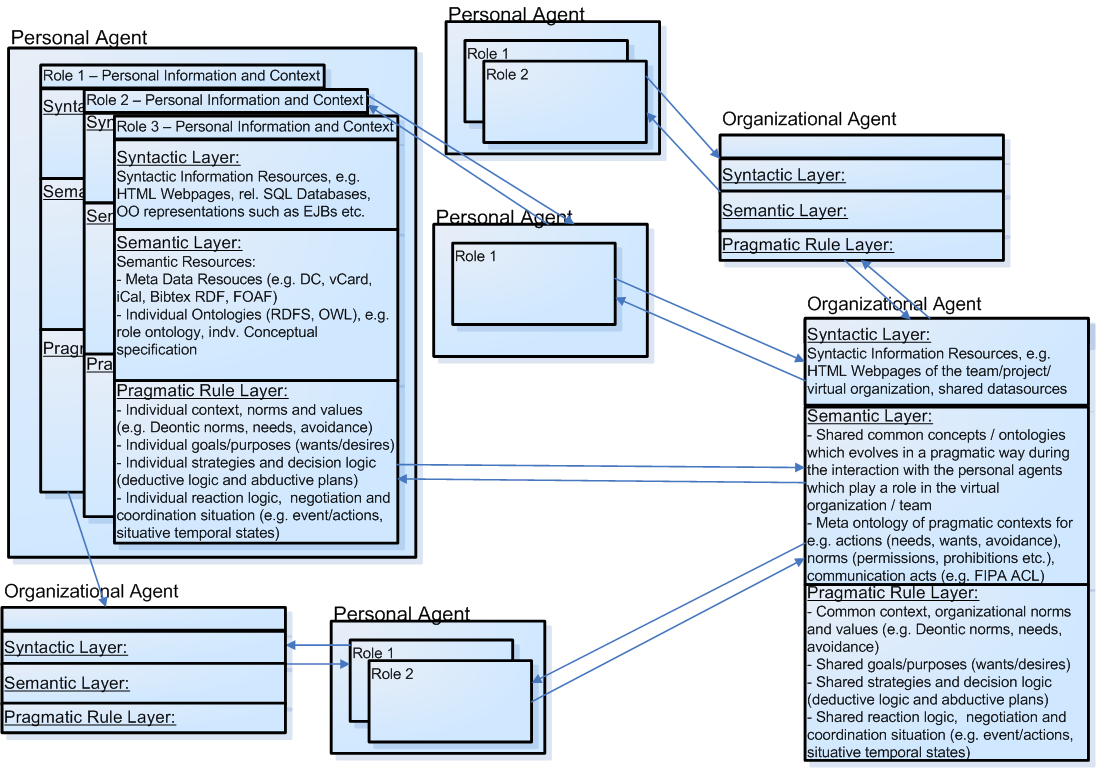}
\caption{A Pragmatic Agent Web for Virtual Organizations}
\label{fig:UseCaseSzenario}
\end{figure}

In the architecture of the eScience Agent Web model(Figure \ref{fig:UseCaseSzenario}), the syntactic level controls the appearance and access of syntactic information resources such as HTML pages. The representation languages such as XML, RDF and OWL on the semantic level make these Web-based resources more readable and processable not only to humans, but also to computers to infer new knowledge. Finally, the pragmatic and behavioral level defines the rules that how information is used and describes the actions in terms of its pragmatic aspects. These rules e.g. transform existing information into relevant information of practical consequences, trigger automated reactions according to occurred complex events, and derive answers  from the existing syntactic and semantic information resources.

In this paper we focus on the pragmatic and behavioral layer and build it upon existing technologies and common language formats of the Semantic Web such as HTML/XML Web pages, RDF/RDFS, OWL and etc. We assume that there is already a critical mass of such data sources on the semantic and syntactic layer. Furthermore, we integrate data and functionality from legacy applications.

\section{Distributed Rule Responder Agent Services}
The core parts of the distributed Rule Responder Architecture for the eScience Agent Web are the common platform-independent rule interchange format (RuleML), the communication middleware (ESB) and the execution environments (Prova).

The Rule Markup Language (RuleML) \cite{Bol06} is a modular, interchangeable rule specification on standard to express both forward and backward rules for deduction, reaction, rewriting, and further inferential-transformational tasks. Reaction RuleML \cite{PKB+06} is a sublanguage of RuleML and incorporates various kinds of production, action, reaction, and KR temporal/event/action logic rules as well as (complex) event/action messages.

To seamlessly handle message-based interactions between the responder agents and with other applications, an enterprise service bus (ESB), the Mule open-source ESB \cite{Mul06} is used. The ESB allows deploying the rule-based agents as highly distributable rule inference services installed as Web-based endpoints in the Mule object broker and supports the Reaction RuleML based communication between them. Mule is based on ideas from ESB architectures, but goes beyond the typical definition of an ESB as a transit system for carrying data between applications.

Prova \cite{KPS06},which is a highly expressive Semantic Web rule engine to the reference implementation for complex agents with complex reaction workflows, decision logic and dynamic access to external Semantic Web data sources.The current version of Prova follows the spirit and design of the recent W3C Semantic Web initiative and combines declarative rules, ontologies and inference with dynamic object-oriented Java API calls and access to external data sources such as relational databases or enterprise applications and IT services.

\section{Rule Responder Use Case}

The discovery process for a researcher to find the Alzheimer's drug target candiates is very complex and time-consuming.He/she first discovers from Uniprot, the W3C HCLS KB and the SWAN data that Beta amyloidal in various forms, and in particular ADDLs, which are good therapeutic targets. He/she then searches the PubMed database about articles on ADDLs and ranks the results to find the top location, which is Evanston, and the top author, who is William Klein. From this, the researcher makes the hypothesis that William Klein works in Evanston, and simply proves it using Google. Finally, the researcher queries the EMBI-EBI database for the patents addressing ADDLs as therapeutic target for AD and concludes that William Klein who also holds two patents is one of the top experts in ADDLs research.Implicitly, the researcher executes the following rule: IF a Person has most publications in the Field and one or more Patents in the field THEN the Person is an expert for this Field. Figure 6 shows how this rule can be implemented in terms of Rule Responder agents.

\begin{figure}
\centering
\includegraphics[height=6.2cm]{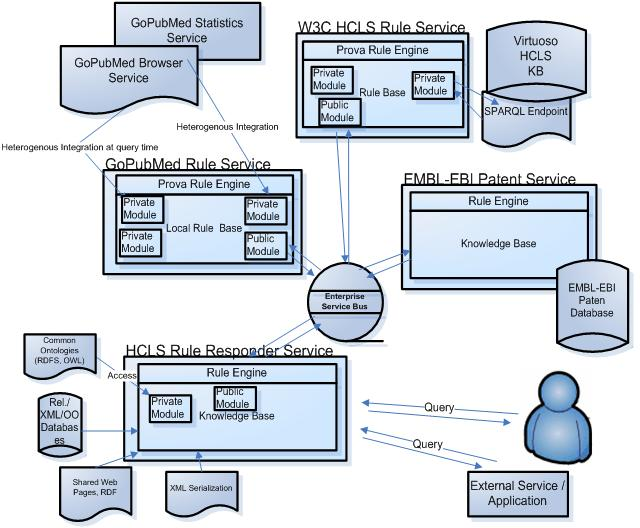}
\caption{Rule Responder HCLS eScience Infrastructure}
\label{fig:AZ_architecture}
\end{figure}

The HCLS Rule Responder agent service (Figure \ref{fig:AZ_architecture}) implements the main logic of the eScience infrastructure and acts as the main communication endpoint for external agents. It’s the rule code defines the public interfaces to receive requests (queries, tasks) to the eScience infrastructure and the logic to look up the respective source agents and delegate requests to them in order to answer the queries and fulfill the tasks. Each existing legacy data sources / service is wrapped by a Rule Responder source agent which runs a Prova rule engine. The agent’s rule base comprises the local rule interface descriptions, i.e. the rule functions which can be queried by other agents of the eScience infrastructure, the respective transformation rules to issue queries to the platform-speciﬁc services and access the heterogeneous local data sources, and the rule logic to process incoming requests and derive answers / information from the local knowledge. 

\section{Conclusion}

With Rule Responder HCLS we have evolved a rule-based approach which facilitates easy heterogeneous systems integration and provides computation, database access, communication, web services, etc. This approach preserves local anonymity of local agent nodes including modularity and information hiding and provides much more control to users with respect to the relatively easy declarative rule-based programming techniques. The rules allow specifying where to access and process information, how to present information and automatically react to it, and how to transform the general information available from existing data sources on the Web into personally relevant information accessible via the eScience infrastructure. The Rule Responder eScience infrastructure is available online at responder.ruleml.org.


\begin{thebibliography}{4}

\bibitem{Bol06} H. Boley. The rule-ml family of web rule languages. In 4th Int. Workshop on Principles and Practice of Semantic Web Reasoning, Budva, Montenegro, 2006.
\bibitem{PKB+06} A. Paschke, et al. Reaction ruleml, http://ibis.in.tum.de/research/reactionruleml/
\bibitem{Mul06}Mule. Mule enterprise service bus, http://mule.codehaus.org/display/mule/home,2006.
\bibitem{KPS06} A. Kozlenkov, et al. Prova, http://prova.ws, 2006.

\end{thebibliography}
\end{document}